# Low-loss $Sb_2S_3$ Optical Phase Shifter Enabled by Optimizing Sputtering Conditions

YUTO MIYATAKE,[1,*] TOMOHIRO ANDA,[1] YOSUKE WAKITA,[1] KOTARO MAKINO,[2] SHOGO HATAYAMA,[2] MAKOTO OKANO,[3] KASIDIT TOPRASERTPONG,[1] SHINICHI TAKAGI,[1] AND MITSURU TAKENAKA[1]

[1] *Department of Electrical Engineering and Information Systems, The University of Tokyo, 7-3-1 Hongo, Bunkyo-ku, Tokyo 113-8656, Japan*
[2] *Semiconductor Frontier Research Center, National Institute of Advanced Industrial Science and Technology (AIST), Tsukuba Central 2, 1-1-1 Umezono, Tsukuba, Ibaraki 305-8568, Japan*
[3] *Photonics-Electronics Integration Research Center, National Institute of Advanced Industrial Science and Technology (AIST), Tsukuba West, 16-1 Onogawa, Tsukuba, Ibaraki 305-8569, Japan*
*miyatake@g.ecc.u-tokyo.ac.jp

**Abstract:** By quantitatively evaluating the atomic concentrations of sputtered $Sb_2S_3$ films with different sputtering powers and Ar flows, we reveal that a sputtered $Sb_2S_3$ film becomes close to the stoichiometric composition as the sputtering power and Ar flow decrease. We characterize the optical properties of $Sb_2S_3$ and show that the lower sputtering power leads to a better figure of merit of $Sb_2S_3$ as an optical phase shifter in the near infrared (NIR) range. Based on these results, we achieve a loss per phase shift of 0.33 dB/π at a wavelength of 1.55 μm, one of the lowest losses among $Sb_2S_3$ phase shifters in the NIR range.

## 1. Introduction

Phase change materials (PCMs), characterized by their non-volatility and a drastic change in optical properties between amorphous and crystalline states, will play a key role in the development of energy-efficient and miniaturized photonic devices and photonic integrated circuits (PICs) [1–3]. Since PCMs exhibit significant changes in the spectra of both the refractive index and extinction coefficient along with the phase change, PCM-based photonic devices and PICs can be divided into two categories: phase-modulation-based ones that use the change in refractive index and intensity-modulation-based ones that use the change in extinction coefficient. Historically, research on intensity-modulation-based photonic devices using $Ge_2Sb_2Te_5$ (GST) has preceded with various demonstrations of GST-based devices including an all-photonic memory [4], non-volatile switches [5–7], a broadband 2×2 switch [8], a plasmonic device with dual electrical-optical functionality[9], etc. Following the research on intensity modulation-based photonic devices, research on phase modulation-based photonic devices began. For the realization of phase modulation-based photonic devices using PCMs, the problem to be solved is the large interband optical absorption in GST at the wavelength of 1.55 μm.

Two different strategies were employed to solve the problem. The first strategy is an approach on the lightwave side, which uses GST in the mid-infrared (MIR) range, taking advantage of the fact that the optical absorption of GST is smaller in the MIR range compared to the near infrared (NIR) range. In [10], the losses of the GST-based phase shifters were dramatically reduced by moving the operating wavelength of the phase modulator from 1.55 μm to 2.32 μm. Low-loss PCM-based phase shifters in the MIR range are promising for optical communications in the 2 μm band, gas sensing, and quantum applications, etc. However, those in the NIR range are more desirable for standard optical communications and optical computing. For this reason, the second material-side strategy is more widely adopted, which attempts to reduce optical absorption at a wavelength of 1.55 μm by using a PCM with a band gap larger than GST. Pioneering work on this strategy was reported in [11], where the extinction coefficients of amorphous and crystalline $Ge_2Sb_2Se_4Te_1$ (GSST), in which Te in GST is partially replaced by Se, are smaller than those of GST. More recently, $Ge_2Sb_2Te_3S_2$ (GSTS), in which Te in GST is partially replaced by S, was proposed and is more suitable for phase shifters than GST [12]. In combination with the first strategy, a low-loss GSTS-based phase shifter has been experimentally demonstrated at a wavelength of 2.34 μm, which is one of the lowest-loss PCM-based phase shifters ever reported.

As wide-gap PCMs for optical phase shifters operating in the NIR range, the new family of wide-gap PCMs, $Sb_2Se_3$ and $Sb_2S_3$, are the most promising PCMs reported so far [13]. This is because these materials are transparent in both the amorphous and crystalline states in the NIR range, while maintaining moderate refractive index contrasts. To date, a number of low-loss optical phase shifters have been successfully demonstrated by using $Sb_2S_3$ [14–19] and $Sb_2Se_3$ [15, 20–25]. $Sb_2S_3$ and $Sb_2Se_3$ are similar to each other with respect to their optical properties, but some differences have been recognized and reported. Compared to $Sb_2S_3$, $Sb_2Se_3$ shows a better switching endurance and reliability, and requires less energy to be switched with a visible laser [13], However, the toxicity of selenium is severe, which makes it difficult for $Sb_2Se_3$ to be introduced in a commercial foundry [17, 26]. This is why we believe $Sb_2S_3$ will have a wider range of applications in the future.

The band gaps of the amorphous and crystalline $Sb_2S_3$ are reported to be about 2.1 eV and 1.7 eV, respectively [27], and at a wavelength of 1.55 μm, corresponding to 0.8 eV, the interband absorption is expected to be zero in both the amorphous and crystalline states. In many reports, the optical absorption of the amorphous state is negligibly small, as expected, while that of the crystalline state is larger than that of the amorphous state, resulting in an optical absorption of more than the order of 0.01 dB/μm in an optical phase shifter. Besides, the extinction coefficients of the crystalline $Sb_2S_3$ and the loss of optical phase shifters using $Sb_2S_3$ have a considerable degree of variation among reports. Although pure refractive index modulation (without extinction coefficient modulation) is desirable for phase-modulation-based photonic devices, neither the cause of the residual extinction coefficient of the crystalline state of $Sb_2S_3$ in the NIR range nor a method to reduce it has been identified.

In this work, we tackle this problem by optimizing the sputtering parameters of $Sb_2S_3$ thin films. In [13], it was reported that the band gap of $Sb_2S_3$ varies significantly depending on the pressure and sputtering power in sputtering, but the relationship between these sputtering parameters and the band gap was not discussed. Inspired by this work, we quantitatively evaluate how the composition of $Sb_2S_3$ depends on the sputtering power and Ar flow during the deposition of $Sb_2S_3$ films. Besides, we characterize the optical properties of the amorphous and crystalline $Sb_2S_3$ deposited with different sputtering powers and reveal that the lower sputtering power leads to a better figure of merit of $Sb_2S_3$ as a phase shifter. Based on these results, we fabricate a low-loss $Sb_2S_3$ phase shifter with the optimized sputtering condition and successfully achieve a loss per phase shift of 0.33 dB/π at a wavelength of 1.55 μm, which is one of the lowest losses among $Sb_2S_3$ phase shifters in the NIR range.

## 2. Deposition and Characterization of $Sb_2S_3$ Films

### 2.1 Preparation of Samples

All $Sb_2S_3$ films used for the following analysis in this section were deposited using a radio-frequency (RF) magnetron sputtering system (QAM-4, ULVAC KYUSYU Corp.) from a $Sb_2S_3$ target (99.9% purity, Kojundo Chemical Laboratory Co., Ltd.) on Si substrates in an argon (Ar) atmosphere. We deposited $Sb_2S_3$ with three different sputtering powers and Ar flows. After depositing the $Sb_2S_3$ film, a $SiO_2$ film was sequentially deposited as a capping layer to prevent the $Sb_2S_3$ from oxidizing.

The as-deposited $Sb_2S_3$ is in its amorphous state. The crystalline $Sb_2S_3$ samples were prepared by heating the as-depo amorphous $Sb_2S_3$ samples on a hot plate at 270 °C for 5 min, and 310 °C for 5 min in the atmosphere. Note that the heating time required for the complete crystallization depends on the deposition conditions and that complete crystallization was achieved for all the deposition conditions by heating at 310 °C after heating at 270 °C.

## 2.2 Analysis of Composition

In order to investigate the composition of $Sb_2S_3$ films with different sputtering powers and Ar flows, we performed Rutherford back-scattering spectrometry (RBS). With RBS analysis, we can calculate the composition ratio of a sample. A collimated 2.275 MeV $4He^{+++}$ ion beam with a beam diameter of 1–2 mm at a backscattering angle of 160° was used. Elements to be detected were O, Si, S, and Sb.

First, we performed RBS analysis for as-depo $Sb_2S_3$ films with different sputtering powers with a fixed Ar flow of 36 sccm. The measured raw RBS spectra are shown in Appendix A. Figure 1 shows the calculated atomic concentration profiles of O, Si, S, and Sb elements. The estimated thicknesses of the top $SiO_2$ film were 11.0 nm, 13.0 nm, and 11.0 nm, and those of $Sb_2S_3$ film were 13.5 nm, 13.5 nm, and 13.0 nm for the sputtering powers of 10 W, 30 W, and 50 W, respectively. The atomic concentrations of Sb and S were (38.8 at.%, 61.2 at.%), (43.2 at.%, 56.8 at.%), and (43.8 at.%, 56.2 at.%) for the sputtering powers of 10 W, 30 W, and 50 W, respectively, which are plotted as a function of the sputtering power in Fig. 3(a). Since the concentrations of Sb to S in the stoichiometric $Sb_2S_3$ are (40.0 at.%, 60.0 at.%), the $Sb_2S_3$ film deposited with 10 W is slightly S-rich, and those with 30 W and 50 W are more pronouncedly Sb-rich. These results show a tendency for the films to become more Sb-rich as the sputtering power increases.

Second, we performed the same analysis for $Sb_2S_3$ films with different Ar flows with a fixed sputtering power of 10 W. The measured raw RBS spectra are also shown in Appendix A. The estimated thicknesses of the top $SiO_2$ film were 11.0 nm, 13.0 nm, and 13.0 nm, and those of $Sb_2S_3$ film were 13.5 nm, 14.0 nm, and 14.0 nm for the sputtering powers of 18 sccm, 36 sccm, and 50 sccm, respectively. The atomic concentrations of Sb and S were (39.4 at.%, 60.6 at.%), (38.8 at.%, 61.2 at.%), and (38.8 at.%, 61.2 at.%) for the sputtering powers of 18 sccm, 36 sccm, and 50 sccm, respectively, which are plotted as a function of the sputtering power in Fig. 3(b). The S and Sb contents approach the stoichiometric ones with decreasing Ar flow. However, compared with the sputtering power dependency, the Ar flow dependency of the atomic concentration is weak. Thus, in the following analysis, we focus on the sputtering power dependence with a fixed Ar flow of 18 sccm, which is expected to produce an $Sb_2S_3$ film with the most stoichiometric S and Sb contents.

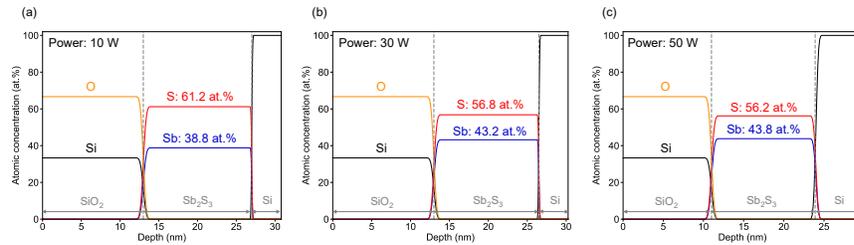

Fig. 1. Atomic concentration profiles of O, Si, S, and Sb elements for the samples with different sputtering powers of (a)10 W, (b) 30 W, and (c) 50 W with an Ar flow of 36 sccm.

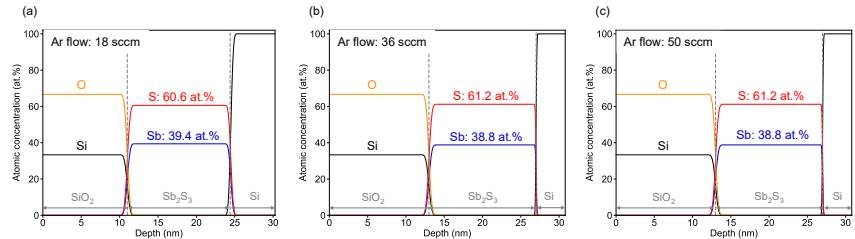

Fig. 2. Atomic concentration profiles of O, Si, S, and Sb elements for the samples with different Ar flows of (a)18 sccm, (b) 36 sccm, and (c) 50 sccm with a sputtering power of 10 W.

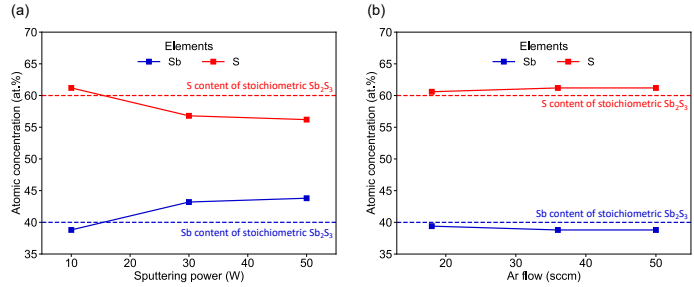

Fig. 3. Atomic concentrations of Sb and S as a function of the sputtering power.

To further characterize the $Sb_2S_3$ films, we measured the Raman scattering spectra of the $Sb_2S_3$ films. We used a Raman microscope (LabRAM HR-800, Horiba, Ltd.) equipped with a 488-nm optically pumped semiconductor laser (Sapphire SF NX, COHERENT). For both amorphous and crystalline $Sb_2S_3$ films, we used 40-mW laser power, 45-sec integration time, 5 accumulations, an optical filter (10% transmission), and a ×100 objective. The laser power and optical filter were determined so as not to induce crystallization due to laser heating.

The measured Raman spectra of $Sb_2S_3$ films sputtered with different sputtering powers with a fixed Ar flow of 18 sccm for the amorphous and crystalline states are shown in Figs. 4(a) and (b), respectively. Amorphous $Sb_2S_3$ films show relatively broad spectra regardless of deposition conditions. In contrast, the spectra of the crystalline states show greatly different shapes depending on the sputtering power. The common peaks exist at 282 and 311 cm$^{-1}$, which are Sb-S stretching modes [28, 29]. The peaks at 113 and 150 cm$^{-1}$ appear as the sputtering power increases. These peaks identify the formation of the Sb-Sb bonding [29, 30]. Therefore, it is implied that the films become Sb-rich as the sputter power increases, and this trend is consistent with the results of the RBS analysis.

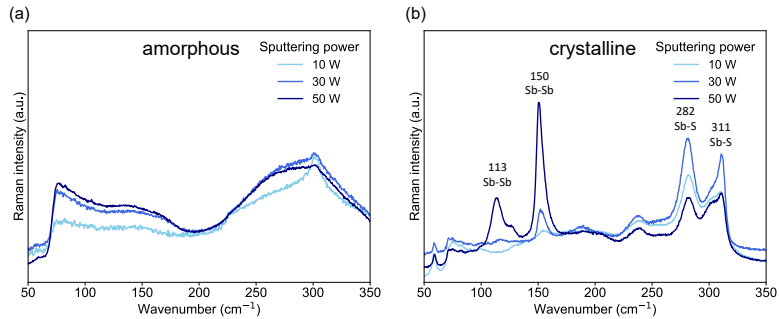

Fig. 4. Raman spectra of (a) amorphous and (b) crystalline $Sb_2S_3$ films sputtered with sputtering powers of 10 W, 30 W, and 50 W with an Ar flow of 18 sccm.

## 2.2 Characterization of Optical Properties

The optical constants of the amorphous and crystalline $Sb_2S_3$ were measured using a spectroscopic ellipsometer (M-2000DI-T, J.A. Woollam Co.) with a measurement wavelength range of 193–1688 nm, an acquisition time of 2 sec, and incident angles of 65°, 70°, and 75°. The measured data was analyzed using the CompleteEASE software in which the data were fitted by an optical model with a three-layer structure of Si/$Sb_2S_3$/$SiO_2$. For the bottom Si and top $SiO_2$ layers, pre-defined "Si_JAW" and "SiO2_JAW2" models were used, respectively. For the $Sb_2S_3$ layer, we used the general oscillator model (Gen-Osc model) with one Tauc–Lorentz oscillator and one Gaussian oscillator. The thicknesses of $SiO_2$ and $Sb_2S_3$ layers were included in the parameters to fit. The fit results are listed in Tables 2 and 3 in Appendix B.

Figures 5(a)–(d) show the refractive index and extinction coefficient spectra of amorphous and crystalline $Sb_2S_3$ sputtered with different sputtering powers from 10 W to 50 W with an Ar flow of 18 sccm. The change in the refractive index spectra of both amorphous and crystalline $Sb_2S_3$ films is not significant with different sputtering powers. The extinction coefficient spectra of both amorphous and crystalline $Sb_2S_3$, however, show a clear trend in which the wavelength at which the extinction coefficient vanishes gets shorter as the sputtering power decreases. For the application to phase shifters, the change in the extinction coefficient of the larger of the two states, i.e., the crystalline state, is important. Although the extinction coefficient of crystalline $Sb_2S_3$ sputtered with 50 W does not disappear even at a wavelength of 1688 nm, those with 30 W and 10 W become zero at wavelengths of 910 nm and 824 nm, respectively. These results indicate that $Sb_2S_3$ deposited with low sputtering powers, such as 10W or 30W, is suitable for realizing low-loss phase shifters. Since the results of the spectroscopic ellipsometry analysis suggest that 10W is expected to be most stoichiometric, we decided to use $Sb_2S_3$ deposited with 10 W for the phase shifters. Note that, in the visible light range, the extinction coefficient of the $Sb_2S_3$ deposited with 10 W is comparable to or higher than others. This is a favorable optical property for amorphization using visible light lasers, for which a large optical absorption in the visible light range is desired for efficient amorphization by an external laser irradiation. The small extinction coefficients of the amorphous $Sb_2S_3$ deposited with 10 W in the visible light range are less problematic because the crystallization temperature is lower than that required for amorphization.

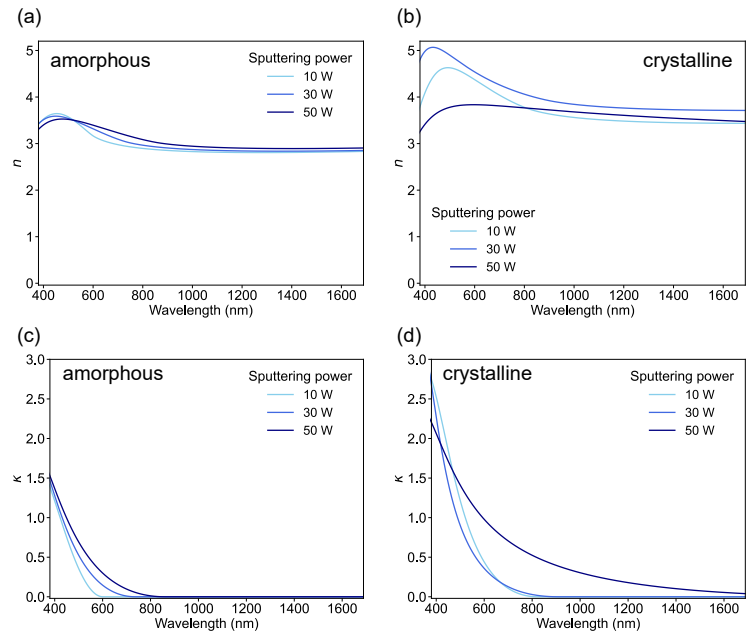

Fig. 5. Refractive index and extinction coefficient spectra of amorphous and crystalline $Sb_2S_3$ sputtered with different sputtering powers of 10 W, 30 W, and 50 W with an Ar flow of 18 sccm. Refractive index spectra of (a) amorphous and (b) crystalline $Sb_2S_3$, and extinction coefficient spectra of (c) amorphous and (d) crystalline $Sb_2S_3$ are plotted.

## 3. Demonstration of Low-loss $Sb_2S_3$ Optical Phase Shifter

*3.1 Fabrication*

To evaluate the loss and phase shift of a $Sb_2S_3$ phase shifter, straight waveguides and micro-ring resonators (MRRs) with different lengths of $Sb_2S_3$ phase shifters were fabricated. The process flow is shown in Fig. 6. The devices were fabricated on a 220-nm silicon-on-insulator (SOI) wafer with a 3-μm thick buried oxide (BOX) layer. Firstly, the Si waveguide patterns were patterned with ArF immersion lithography, followed by full etching of Si using dry etching. Then focusing grating couplers (GCs) were patterned with KrF lithography and partial etching of Si using dry etching with an etching depth of 140 nm. The pitch and filling factor of the uniform grating were 688 nm and 0.6, respectively. After the patterning of the SOI layer, 500-nm-thick $SiO_2$ was deposited using high-density plasma chemical vapor deposition (HDP-CVD), which was thinned down to around 30–50 nm using chemical mechanical polishing (CMP) and dilute hydrogen fluoride (DHF) wet etching. Note that this process of forming embedded $SiO_2$ was for other devices fabricated on the same wafer and is not necessarily essential for the fabrication of $Sb_2S_3$ phase shifters.

As a cladding layer, 400-nm thick $SiO_2$ was deposited using plasma-enhanced chemical vapor deposition (PECVD). To deposit $Sb_2S_3$ film on top of the Si waveguide, 2-μm wide windows were opened into the $SiO_2$ cladding using electron-beam (EB) lithography, followed by inductively coupled plasma reactive ion etching (ICP-RIE) with Ar and $C_4F_8$ and buffered hydrogen fluoride (BHF) wet etching. Afterwards, 20-nm thick $Sb_2S_3$ and 20-nm thick $SiO_2$ films were deposited sequentially using radiofrequency (RF) sputtering. Finally, the $Sb_2S_3$ and $SiO_2$ films were patterned except for the phase shifter region by EB lithography and ICP-RIE with Ar&$C_4F_8$ (for $SiO_2$) and Ar (for $Sb_2S_3$). In order to evaluate the loss and phase shift per unit length of $Sb_2S_3$ phase shifter, we prepared straight waveguides and micro-ring resonators with different lengths of $Sb_2S_3$ phase shifters as shown in Fig. 7.

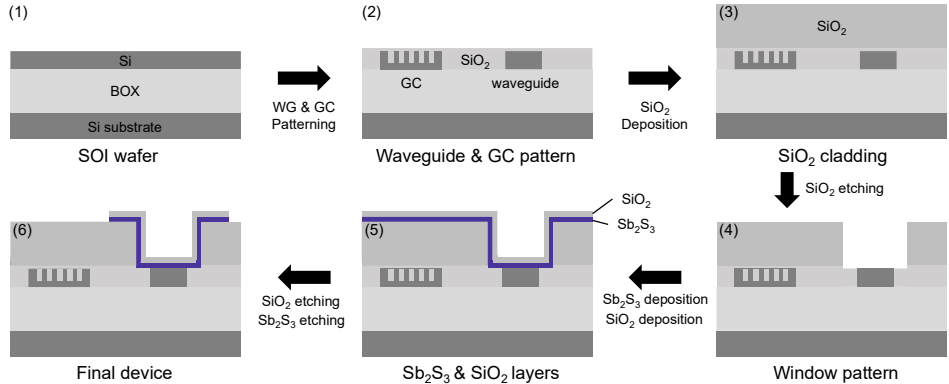

Fig. 6. Process flow of $Sb_2S_3$ phase shifter.

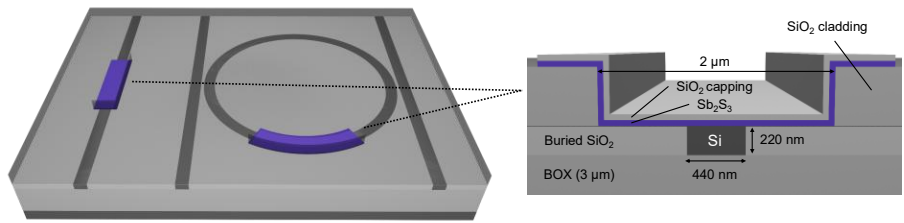

Fig. 7. Three-dimensional and cross-sectional view of a straight waveguide and MRR with a $Sb_2S_3$ phase shifter.

## 3.2 Measurement Setup

Figure 8 shows a schematic of the measurement setup. The input light was supplied by a tunable laser source (TSL-710, Santec) via a single-mode fiber (SMF), and its polarization was controlled by a manual fiber polarization controller to match the fundamental TE mode of the waveguides. Focusing GCs were used to couple light into and out of the devices under test at an angle of ~10° using optical probes (Light Wave Probe, FormFactor, Inc.). An optical power meter (MPM-210H, Santec) was used to measure optical transmission. By connecting the tunable laser source and optical power meter with two BNC cables and controlling these devices with a computer via GPIB and USB cables, a high-speed wavelength sweeping system was built.

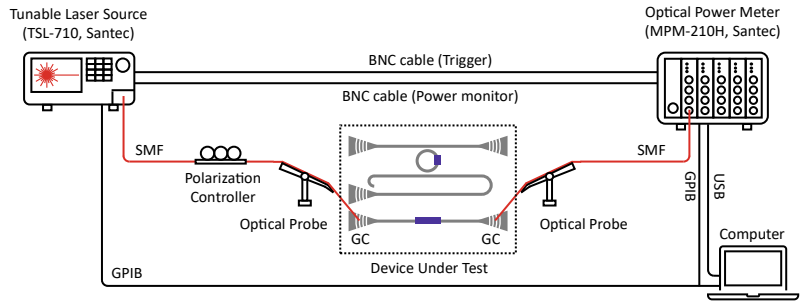

Fig. 8. Schematic of the measurement setup.

## 3.2 Measurement of Loss

The loss per unit length of the $Sb_2S_3$ phase shifter was measured using straight waveguides with different lengths of the phase shifter, as shown in Fig 8(a). Figures 8(b) and (c) show the transmission spectra of the straight waveguides with different lengths of the $Sb_2S_3$ phase shifters up to 100 μm after annealing the sample at 110 °C and 310 °C for 5 min, respectively. The wavelength dependence of the transmission spectrum is attributed to that of the GC. Note that to ensure the non-volatile operation of the phase shifters, all spectra were measured at room temperature after the chip had cooled down.

Figure 9(d) shows the calculated insertion loss at a wavelength of 1.55 μm as a function of the phase shifter length at different annealing temperatures. The extracted losses per unit length at a wavelength of 1.55 μm with an annealing temperature of 110 °C and 310 °C are 0.007 dB/μm and 0.013 dB/μm, respectively. According to the spectroscopic ellipsometry measurement results, the extinction coefficients of both amorphous and crystalline $Sb_2S_3$ are zero, so the loss per unit length of the phase shifter would ideally be equal to the propagation loss of the Si strip waveguide, which is expected to be ~ $1.5 \times 10^{-4}$ dB/μm. The excess loss of the phase shifter with amorphous $Sb_2S_3$ would mainly come from light scattering due to the roughness of the $Sb_2S_3$ and $SiO_2$ films. For the phase shifter with crystalline $Sb_2S_3$ light scattering at grain boundaries would have further increased losses.

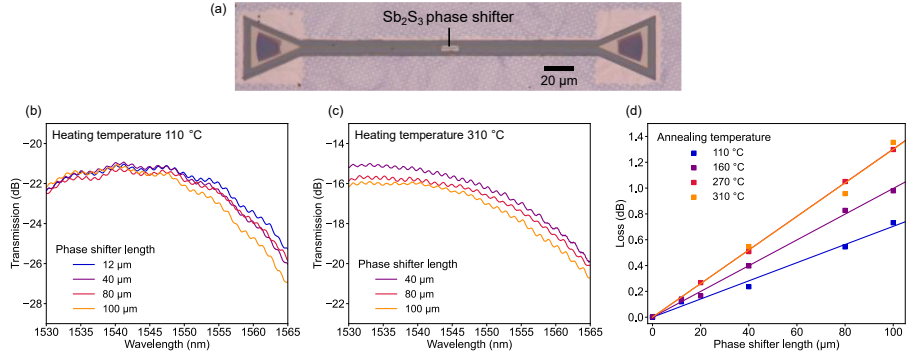

Fig. 9. (a) Optical microscopy image of a straight waveguide with a $Sb_2S_3$ phase shifter. Transmission spectra of straight waveguides with different lengths of the $Sb_2S_3$ phase shifters after annealing the sample at (a) 110 °C and (b) 310 °C. (d) Relationship between the optical loss and length of the $Sb_2S_3$ phase shifter for a 1.55-µm wavelength for annealing temperatures from 110 °C to 310 °C.

*3.3 Measurement of Phase Shift*

The phase shift per unit length of the $Sb_2S_3$ phase shifter was measured using add-drop MRRs with different lengths of the phase shifter, as shown in Fig 10(a). The radius and gap of MRR were 20 µm and 200 nm, respectively, and the free spectral range (FSR) around a wavelength of 1.55 µm was 4.4 nm. Figures 10(b) and (c) show the normalized transmission spectra to the bar port of an add-drop MRR with a 10-µm and 20-µm long $Sb_2S_3$ phase shifter measured after annealing the sample at different temperatures from 110 °C to 310 °C. As the refractive index change along with the phase change of $Sb_2S_3$ is positive, the resonance wavelengths shift towards the longer wavelengths. Even after crystallization, the extinction ratio (ER) was almost unchanged compared to before the crystallization, reflecting the very low optical absorption of the crystalline $Sb_2S_3$.

Figure 10(d) shows the calculated phase shift at a wavelength of 1.55 µm as a function of the phase shifter length at different annealing temperatures. The extracted phase shifts per unit length at a wavelength of 1.55 µm with an annealing temperature of 310 °C are 0.041 $\pi$/µm, which corresponds to 24-µm-long phase shifter for a $\pi$ phase shift. This measured loss per unit length is smaller than the simulated one (0.067 $\pi$/µm), which was obtained by numerical analysis for a phase shifter with 20-nm thick $Sb_2S_3$ using an eigenmode solver based on the spectroscopic ellipsometry measurement results. This discrepancy would be because the actual $Sb_2S_3$ film thickness was thinner than the target of 20 nm. If we assume a $Sb_2S_3$ film thickness of about 12 nm, the reduced phase shift is justified. Actually, the assumed film thickness agrees well with the value obtained from the RBS analysis.

Combined with the result of the loss measurement, the loss per $\pi$ phase shift of the $Sb_2S_3$ phase shifter is 0.33 dB/$\pi$ at a wavelength of 1.55 µm. This result is benchmarked in the next section.

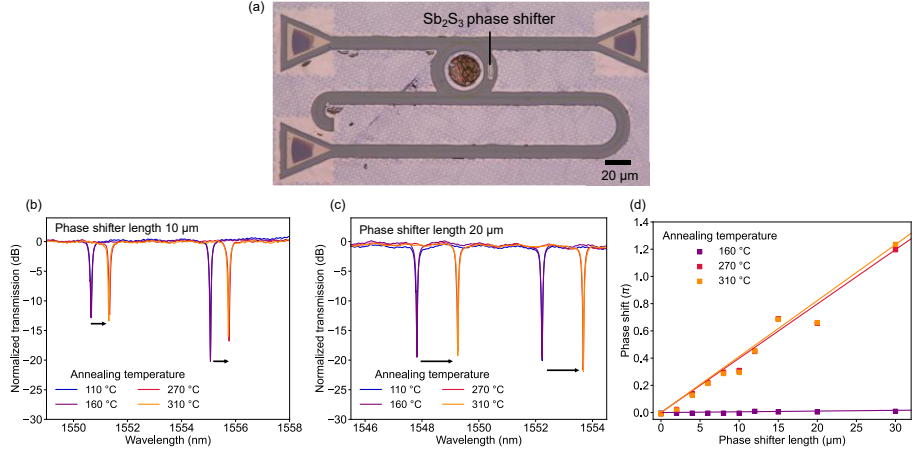

Fig. 10. (a) Optical microscopy image of an add-drop MRR with a $Sb_2S_3$ phase shifter. Normalized transmission spectra to the bar port of an add-drop of an MRR with (d) 10-μm-long and (c) 20-μm-long $Sb_2S_3$ phase shifter measured after annealing the sample at different temperatures. (d) Relationship between the phase shift and length of the $Sb_2S_3$ phase shifter for a 1.55-μm wavelength for annealing temperatures from 160 °C to 310 °C.

### 3.4 Benchmark

Figure 11 shows the experimental results of losses per phase shift of $Sb_2S_3$ phase shifters as a function of the sputtering power. The data plotted with a circle marker were obtained from measurements of a phase shifter, while those plotted with a square marker were calculated from the ellipsometry measurements based on the equations (1) and (2) in [12]. The sputtering powers reported in [14], [31], and [32] are 27 sccm, 20 sccm, and 30 sccm, respectively. Since the data has been collected from different groups who may use different machines to sputter $Sb_2S_3$, the variation in the losses is not small. However, the set of data seems to follow a global trend that the optical loss of a sputtered $Sb_2S_3$ film decreases as the sputtering power decreases.

In Fig. 12, the losses per phase shift of experimentally demonstrated PCM-based optical phase shifters are plotted as a function of their operating wavelength. The value of the loss per phase shift, as well as the loss per unit length and phase shift per unit length, are summarized in Table I. Four different PCMs have been used for these PCM-based phase shifters: GST, GSTS, $Sb_2Se_3$, and $Sb_2S_3$. Among these PCMs, only $Sb_2Se_3$ suffers from the toxicity of Se. As mentioned earlier, the loss per phase shift of the GST phase shifter (blue dots) can be improved in the MIR range compared to those in the NIR range, but the loss is still relatively high even in the MIR range. This problem is mitigated by using GSTS (purple dots) instead of GST, resulting in 0.29 dB/π at a wavelength of 2.34 μm in the lower right corner, which is the lowest loss per phase shift of the Se-free PCM-based phase shifter. However, at a wavelength of 1.55 μm in the NIR range, the loss remains significant.

By using $Sb_2Se_3$ and $Sb_2S_3$, low-loss PCM-based phase shifters have been realized at wavelengths of 1.31 and 1.55 μm (orange and red dots, respectively). The long major axis of the red ellipse reflects the large variation in the loss of phase shifters using these PCMs. The $Sb_2S_3$ phase shifter demonstrated in this work shows the loss per phase shift of 0.33 dB/π at a wavelength of 1.55 μm, which is one of the lowest losses among $Sb_2S_3$ phase shifters, in other words, among Se-free PCM-based phase shifters, in the NIR range.

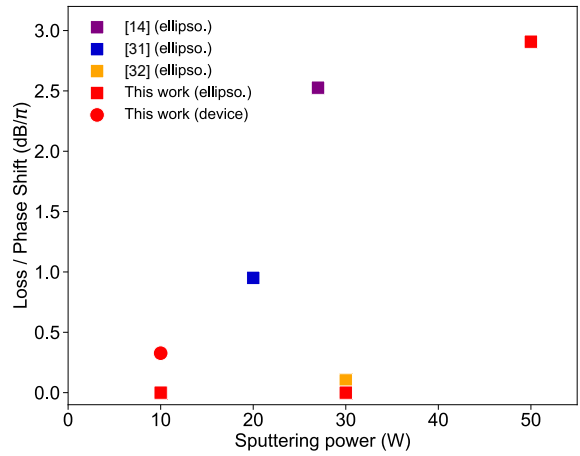

Fig. 11. Experimental results of losses per phase shift of $Sb_2S_3$ phase shifters as a function of the sputtering power.

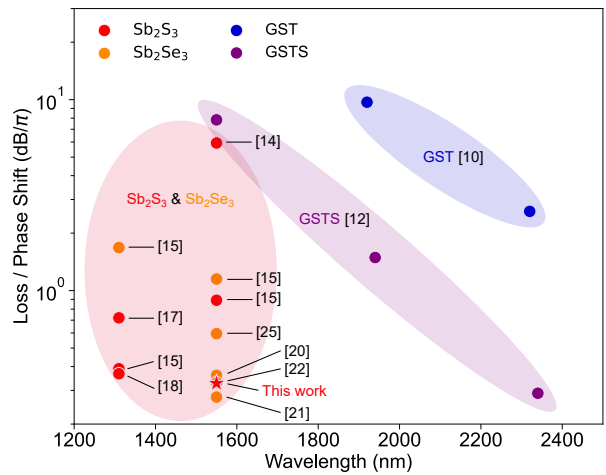

Fig. 12. Loss per phase shift of experimentally demonstrated low-loss optical phase shifter based on PCMs.

Table 1. Comparison of Low-loss Optical Phase Shifter based on PCMs

| PCM | Toxicity of Se | Reference | Wavelength (nm) | Loss[a] (dB/μm) | Phase Shift (π/μm) | Loss per Phase Shift (dB/π) |
|---|---|---|---|---|---|---|
| GST | Free | [10] | 1920 | 0.251 | 0.025 | 9.72 |
|  |  | [10] | 2320 | 0.113 | 0.040 | 2.62 |
| GSTS | Free | [12] | 1550 | 1.700 | 0.216 | 7.84 |
|  |  | [12] | 1940 | 0.199 | 0.124 | 1.49 |
|  |  | [12] | 2340 | 0.026 | 0.089 | 0.29 |
| $Sb_2Se_3$ | Contained | [15] | 1310 | 0.077 | 0.046 | 1.68 |
|  |  | [15] | 1550 | 0.070 | 0.065 | 1.15 |
|  |  | [20] | 1550 | 0.030 | 0.083 | 0.36 |
|  |  | [21] | 1550 | 0.019 | 0.069 | 0.28 |
|  |  | [22] | 1550 | 0.030 | 0.090 | 0.33 |
|  |  | [25] | 1550 | 0.022 | 0.038 | 0.60 |
| $Sb_2S_3$ | Free | [14] | 1550 | 0.160 | 0.270 | 5.93 |
|  |  | [15] | 1310 | 0.031 | 0.076 | 0.39 |
|  |  | [15] | 1550 | 0.023 | 0.026 | 0.89 |
|  |  | [17] | 1310 | 0.024 | 0.033 | 0.72 |
|  |  | [18] | 1310 | 0.004 | 0.012 | 0.37 |
|  |  | **This work** | **1550** | **0.013** | **0.041** | **0.33** |

[a]Loss of optical phase shifter with crystalline PCM

## 4. Conclusions

In this paper, we have experimentally demonstrated a low-loss $Sb_2S_3$ phase shifter by optimizing the sputtering conditions. We have quantitatively evaluated the atomic concentrations of sputtered $Sb_2S_3$ films with different sputtering powers and Ar flows, revealing that a sputtered $Sb_2S_3$ film becomes close to the stoichiometric composition as the sputtering power and Ar flow decreases. In addition, we have characterized the optical properties of the amorphous and crystalline $Sb_2S_3$ deposited with different sputtering powers and have shown that the lower sputtering power leads to a better figure of merit of $Sb_2S_3$ as a phase shifter. Based on these results, we have fabricated a low-loss $Sb_2S_3$ phase shifter with the optimized sputtering condition. We have evaluated the loss and phase shift per unit length of the $Sb_2S_3$ phase shifter using straight waveguides and MRRs and have successfully achieved a loss per phase shift of 0.33 dB/π at a wavelength of 1.55 μm, which is one of the lowest losses among $Sb_2S_3$ phase shifters in the NIR range.

The demonstrated $Sb_2S_3$ phase shifter is highly attractive as a compact, Se-free, low-loss, and non-volatile optical phase shifter, which will be a key enabler for scaling up photonic integrated circuits. Additionally, the obtained low-loss $Sb_2S_3$ film with the optimized sputtering conditions will be an excellent "canvas" on which arbitrary low-loss optical components can be written with a laser beam in the manner reported in [33–35]. We believe that achievement in this work will fundamentally improve the potential capabilities of rapidly expanding PCM-based photonics.

## Appendix A: Fit results of spectroscopic ellipsometry analysis

The measured raw RBS analysis for as-depo $Sb_2S_3$ films with different sputtering powers with a fixed Ar flow of 36 sccm are shown in Fig. 13. Similarly, the measured raw RBS analysis for as-depo $Sb_2S_3$ films with different Ar flows with a fixed sputtering power of 10 W are shown in Fig. 14. The peaks around channel numbers 305 and 449 correspond to the signal from S and

Sb, respectively. The composition ratio can be determined from the ratio of the areas of these peaks.

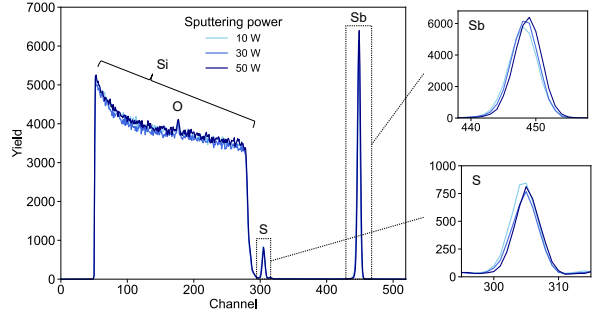

Fig. 13. RBS spectra of $Sb_2S_3$ films sputtered with different sputtering powers of 10 W, 30 W, and 50 W with an Ar flow of 18 sccm.

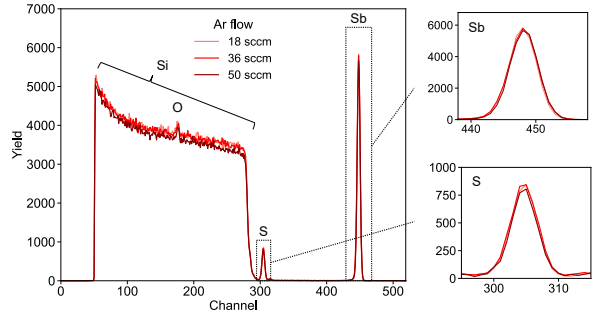

Fig. 14. RBS spectra of $Sb_2S_3$ films sputtered with different Ar flows of 18 sccm, 36 sccm and 50 sccm with a sputtering power of 10 W.

## Appendix B: Fit results of spectroscopic ellipsometry analysis

The optical constants of the amorphous and crystalline $Sb_2S_3$ were measured using a spectroscopic ellipsometry. The mean squared error (MSE) of the fitting and the fitted thicknesses of $SiO_2$ and $Sb_2S_3$ are listed in Tables 2 and 3.

Table 2. Fit results of amorphous $Sb_2S_3$ with different sputtering powers

| Parameter | Sputtering Power: 10 W | Sputtering Power: 30 W | Sputtering Power: 50 W |
| --- | --- | --- | --- |
| MSE | 7.467 | 9.784 | 8.102 |
| Thickness $SiO_2$ | 17.28 ± 0.065 nm | 16.46 ± 0.083 nm | 15.71 ± 0.069 nm |
| Thickness $Sb_2S_3$ | 20.44 ± 0.051 nm | 20.00 ± 0.087 nm | 22.89 ± 0.081 nm |

Table 3. Fit results of crystalline $Sb_2S_3$ with different sputtering powers

| Parameter | Sputtering Power: 10 W | Sputtering Power: 30 W | Sputtering Power: 50 W |
| --- | --- | --- | --- |
| MSE | 6.767 | 11.537 | 10.836 |
| Thickness $SiO_2$ | 15.92 ± 0.048 nm | 19.33 ± 0.066 nm | 15.31 ± 0.095 nm |
| Thickness $Sb_2S_3$ | 18.13 ± 0.124 nm | 18.14 ± 0.162 nm | 19.33 ± 0.148 nm |

**Funding.** Japan Society for the Promotion of Science (JP20H0219); Japan Science and Technology Agency (JPMJCR2004); Ministry of Education, Culture, Sports, Science and Technology (JPMXP1224UT0063, JPMXP1224UT1028).

**Acknowledgment.** A part of this work was supported by "Advanced Research Infrastructure for Materials and Nanotechnology in Japan (ARIM)" of the Ministry of Education, Culture, Sports, Science and Technology (MEXT), Grant Number JPMXP1224UT0063. We acknowledge J. Tominaga and Y. Saito for their support in the early stage of this work.

**Disclosures.** The authors declare no conflicts of interest.

### References

1. M. Wuttig, H. Bhaskaran, and T. Taubner, "Phase-change materials for non-volatile photonic applications," Nat. Photonics **11**(8), 465–476 (2017).
2. M. S. Nisar, X. Yang, L. Lu, J. Chen, and L. Zhou, "On-chip integrated photonic devices based on phase change materials," Photonics **8**(6), 205 (2021).
3. P. Prabhathan, K. V. Sreekanth, J. Teng, J. H. Ko, Y. J. Yoo, H.-H. Jeong, Y. Lee, S. Zhang, T. Cao, C.-C. Popescu, B. Mills, T. Gu, Z. Fang, R. Chen, H. Tong, Y. Wang, Q. He, Y. Lu, Z. Liu, H. Yu, A. Mandal, Y. Cui, A. S. Ansari, V. Bhingardive, M. Kang, C. K. Lai, M. Merklein, M. J. Müller, Y. M. Song, Z. Tian, J. Hu, M. Losurdo, A. Majumdar, X. Miao, X. Chen, B. Gholipour, K. A. Richardson, B. J. Eggleton, K. Sharda, M. Wuttig, and R. Singh, "Roadmap for phase change materials in photonics and beyond," iScience **26**(10), 107946 (2023).
4. C. Ríos, M. Stegmaier, P. Hosseini, D. Wang, T. Scherer, C. D. Wright, H. Bhaskaran, and W. H. P. Pernice, "Integrated all-photonic non-volatile multi-level memory," Nat. Photonics **9**(11), 725–732 (2015).
5. Y. Ikuma, Y. Shoji, M. Kuwahara, X. Wang, K. Kintaka, H. Kawashima, D. Tanaka, and H. Tsuda, "Small-sized optical gate switch using Ge2Sb2Te5 phase-change material integrated with silicon waveguide," Electron. Lett. **46**(5), 368 (2010).
6. H. Zhang, L. Zhou, L. Lu, J. Xu, N. Wang, H. Hu, B. M. A. Rahman, Z. Zhou, and J. Chen, "Miniature multilevel optical memristive switch using phase change material," ACS Photonics **6**(9), 2205–2212 (2019).
7. J. Zheng, Z. Fang, C. Wu, S. Zhu, P. Xu, J. K. Doylend, S. Deshmukh, E. Pop, S. Dunham, M. Li, and A. Majumdar, "Nonvolatile Electrically Reconfigurable Integrated Photonic Switch Enabled by a Silicon PIN Diode Heater," Adv. Mater. **32**(31), e2001218 (2020).
8. R. Chen, Z. Fang, J. E. Fröch, P. Xu, J. Zheng, and A. Majumdar, "Broadband nonvolatile electrically controlled programmable units in silicon photonics," ACS Photonics **9**(6), 2142–2150 (2022).
9. N. Farmakidis, N. Youngblood, X. Li, J. Tan, J. L. Swett, Z. Cheng, C. D. Wright, W. H. P. Pernice, and H. Bhaskaran, "Plasmonic nanogap enhanced phase-change devices with dual electrical-optical functionality," Sci. Adv. **5**(11), eaaw2687 (2019).
10. Y. Miyatake, C. P. Ho, P. Pitchappa, R. Singh, K. Makino, J. Tominaga, N. Miyata, T. Nakano, K. Toprasertpong, S. Takagi, and M. Takenaka, "Non-volatile compact optical phase shifter based on $Ge_2Sb_2Te_5$ operating at 2.3 μm," Opt. Mater. Express **12**(12), 4582 (2022).
11. Y. Zhang, J. B. Chou, J. Li, H. Li, Q. Du, A. Yadav, S. Zhou, M. Y. Shalaginov, Z. Fang, H. Zhong, C. Roberts, P. Robinson, B. Bohlin, C. Ríos, H. Lin, M. Kang, T. Gu, J. Warner, V. Liberman, K. Richardson, and J. Hu, "Broadband transparent optical phase change materials for high-performance nonvolatile photonics," Nat. Commun. **10**(1), 4279 (2019).
12. Y. Miyatake, K. Makino, J. Tominaga, N. Miyata, T. Nakano, M. Okano, K. Toprasertpong, S. Takagi, and M. Takenaka, "Proposal of low-loss non-volatile mid-infrared optical phase shifter based on $Ge_2Sb_2Te_3S_2$," IEEE Trans. Electron Devices **70**(4), 2106–2112 (2023).
13. M. Delaney, I. Zeimpekis, D. Lawson, D. W. Hewak, and O. L. Muskens, "A new family of ultralow loss reversible phase‐change materials for photonic integrated circuits: Sb 2 S 3 and Sb 2 Se 3," Adv. Funct. Mater. **30**(36), 2002447 (2020).
14. Z. Fang, J. Zheng, A. Saxena, J. Whitehead, Y. Chen, and A. Majumdar, "Non‐volatile reconfigurable integrated photonics enabled by broadband low‐loss phase change material," Adv. Opt. Mater. **9**(9), 2002049 (2021).
15. J. Faneca, I. Zeimpekis, S. T. Ilie, T. D. Bucio, K. Grabska, D. W. Hewak, and F. Y. Gardes, "Towards low loss non-volatile phase change materials in mid index waveguides," Neuromorph. Comput. Eng. **1**(1), 014004 (2021).
16. L. Lu, S. F. G. Reniers, Y. Wang, Y. Jiao, and R. E. Simpson, "Reconfigurable InP waveguide components using the $Sb_2S_3$ phase change material," J. Opt. **24**(9), 094001 (2022).
17. R. Chen, Z. Fang, C. Perez, F. Miller, K. Kumari, A. Saxena, J. Zheng, S. J. Geiger, K. E. Goodson, and A. Majumdar, "Non-volatile electrically programmable integrated photonics with a 5-bit operation," Nat. Commun. **14**(1), 3465 (2023).
18. R. Chen, V. Tara, M. Choi, J. Dutta, J. Sim, J. Ye, Z. Fang, J. Zheng, and A. Majumdar, "Deterministic quasi-continuous tuning of phase-change material integrated on a high-volume 300-mm silicon photonics platform," npj Nanophotonics **1**(1), 1–9 (2024).
19. A. Biegański, M. Perestjuk, R. Armand, A. Della Torre, C. Laprais, G. Saint-Girons, V. Reboud, J.-M. Hartmann, J.-H. Tortai, A. Moreau, J. Lumeau, T. Nguyen, A. Mitchell, C. Monat, S. Cueff, and C. Grillet, "$Sb_2S_3$ as a low-loss phase-change material for mid-IR photonics," Opt. Mater. Express **14**(4), 862 (2024).

20. C. Ríos, Y. Zhang, Q. Du, C.-C. Popescu, M. Shalaginov, P. Miller, C. M. Roberts, M. Kang, K. A. Richardson, S. An, C. Fowler, H. Zhang, T. Gu, S. A. Vitale, and J. Hu, "Electrically-switchable foundry-processed phase change photonic devices," in *Active Photonic Platforms XIII*, G. S. Subramania and S. Foteinopoulou, eds. (SPIE, 2021), **11796**, pp. 51–58.
21. K. Lei, M. Wei, Z. Chen, J. Wu, J. Jian, J. Du, J. Li, L. Li, and H. Lin, "Magnetron-sputtered and thermal-evaporated low-loss Sb-Se phase-change films in non-volatile integrated photonics," Opt. Mater. Express **12**(7), 2815 (2022).
22. C. Ríos, Q. Du, Y. Zhang, C.-C. Popescu, M. Y. Shalaginov, P. Miller, C. Roberts, M. Kang, K. A. Richardson, T. Gu, S. A. Vitale, and J. Hu, "Ultra-compact nonvolatile phase shifter based on electrically reprogrammable transparent phase change materials," PhotoniX **3**(1), 1–13 (2022).
23. Z. Fang, R. Chen, J. Zheng, A. I. Khan, K. M. Neilson, S. J. Geiger, D. M. Callahan, M. G. Moebius, A. Saxena, M. E. Chen, C. Rios, J. Hu, E. Pop, and A. Majumdar, "Ultra-low-energy programmable non-volatile silicon photonics based on phase-change materials with graphene heaters," Nat. Nanotechnol. **17**(8), 842–848 (2022).
24. X. Yang, L. Lu, Y. Li, Y. Wu, Z. Li, J. Chen, and L. Zhou, "Non-Volatile Optical Switch Element Enabled by Low-Loss Phase Change Material," Adv. Funct. Mater. **33**(42), 2304601 (2023).
25. M. Wei, K. Xu, B. Tang, J. Li, Y. Yun, P. Zhang, Y. Wu, K. Bao, K. Lei, Z. Chen, H. Ma, C. Sun, R. Liu, M. Li, L. Li, and H. Lin, "Monolithic back-end-of-line integration of phase change materials into foundry-manufactured silicon photonics," Nat. Commun. **15**(1), 2786 (2024).
26. J. E. Spallholz, "On the nature of selenium toxicity and carcinostatic activity," Free Radic. Biol. Med. **17**(1), 45–64 (1994).
27. W. Dong, H. Liu, J. K. Behera, L. Lu, R. J. H. Ng, K. V. Sreekanth, X. Zhou, J. K. W. Yang, and R. E. Simpson, "Wide bandgap phase change material tuned visible photonics," Adv. Funct. Mater. **29**(6), 1806181 (2019).
28. P. Sereni, M. Musso, P. Knoll, P. Blaha, K. Schwarz, G. Schmidt, P. M. Champion, and L. D. Ziegler, "Polarization-dependent Raman characterization of Stibnite (Sb[sub 2]S[sub 3])," in *AIP Conference Proceedings* (AIP, 2010), **1267**, pp. 1131–1132.
29. R. Parize, T. Cossuet, O. Chaix-Pluchery, H. Roussel, E. Appert, and V. Consonni, "In situ analysis of the crystallization process of Sb 2 S 3 thin films by Raman scattering and X-ray diffraction," Mater. Des. **121**, 1–10 (2017).
30. Policies & Practices and P. Conduct, "Effect of pressure on the Raman modes of antimony," (2006).
31. T. Y. Teo, M. Krbal, J. Mistrik, J. Prikryl, L. Lu, and R. E. Simpson, "Comparison and analysis of phase change materials-based reconfigurable silicon photonic directional couplers," Opt. Mater. Express **12**(2), 606 (2022).
32. T. Y. Teo, N. Li, L. Y. M. Tobing, A. S. K. Tong, D. K. T. Ng, Z. Ren, C. Lee, L. Y. T. Lee, and R. E. Simpson, "Capping layer effects on $Sb_2S_3$-based reconfigurable photonic devices," ACS Photonics **10**(9), 3203–3214 (2023).
33. H. Liu, W. Dong, H. Wang, L. Lu, Q. Ruan, Y. S. Tan, R. E. Simpson, and J. K. W. Yang, "Rewritable color nanoprints in antimony trisulfide films," Sci. Adv. **6**(51), eabb7171 (2020).
34. F. Miller, R. Chen, J. Fröch, Z. Fang, V. Tara, S. Geiger, and A. Majumdar, "Rewritable photonic integrated circuit canvas based on low-loss phase change material and nanosecond pulsed lasers," Nano Lett. **24**(23), 6844–6849 (2024).
35. G. Shixin, R. Haonan, P. Jingzhe, R. Yang, Z. Zhenqing, Z. Shuang, and C. Tun, "Reconfigurable free form silicon photonics by phase change waveguides," Adv. Opt. Mater. 2402997 (2025).